\documentclass[aps,pra,twocolumn,floatfix,longbibliography, superscriptaddress]{revtex4-1}

\usepackage{xcolor}
	
\usepackage{graphicx}

\usepackage{pifont}

\usepackage{float}
\usepackage{epsfig}
\usepackage{bm}
\usepackage[T1]{fontenc}
\usepackage[english]{babel}
\usepackage{amsmath}
\usepackage{amssymb}
\usepackage{color}
\usepackage{graphicx}
\usepackage{bbm}
\usepackage{url}
\usepackage{nomencl}
\usepackage{subfigure}
\usepackage{slashed}

\newcommand{\ket}[1]{|#1\rangle}

\newcommand{\figref}[1]{\mbox{Fig.~\ref{#1}}}

\renewcommand{\eqref}[1]{\mbox{Eq.~(\ref{#1})}}

\newcommand{\be}{\begin{equation}}
\newcommand{\ee}{\end{equation}}
\newcommand{\bea}{\begin{eqnarray}}
\newcommand{\eea}{\end{eqnarray}}
\newcommand{\beal}{\begin{align}}
\newcommand{\eeal}{\end{align}}
\usepackage{xr}
\usepackage[colorlinks]{hyperref}
\hypersetup{%
    plainpages=true,
    breaklinks=true,
    hypertexnames=false,
    pageanchor=true,
    colorlinks=true,
    linkcolor={blue},
    citecolor={red},
    urlcolor={blue},
    anchorcolor={black}
}

\makeatletter

\begin{document}

	\title{Gauge Principle and Gauge Invariance in Quantum Two-Level Systems}

\author{Salvatore Savasta}
\email[corresponding author: ]{ssavasta@unime.it}
\affiliation{Dipartimento di Scienze Matematiche e Informatiche, Scienze Fisiche e  Scienze della Terra,
	Universit\`{a} di Messina, I-98166 Messina, Italy}

\author{Omar Di Stefano}
\affiliation{Dipartimento di Scienze Matematiche e Informatiche, Scienze Fisiche e  Scienze della Terra, Universit\`{a} di Messina, I-98166 Messina, Italy}	

\author{Alessio Settineri}
	\affiliation{Dipartimento di Scienze Matematiche e Informatiche, Scienze Fisiche e  Scienze della Terra, Universit\`{a} di Messina, I-98166 Messina, Italy}
	
\author{David Zueco}
	\affiliation {Instituto de Ciencia de Materiales de
		Arag\'{o}n and Departamento de F\'{i}sica de la Materia Condensada ,
		CSIC-Universidad de Zaragoza, Pedro Cerbuna 12, 50009 Zaragoza,
		Spain}
	\affiliation{Fundaci\'{o}n ARAID, Campus R\'{i}o Ebro, 50018 Zaragoza, Spain}
	\author{Stephen Hughes}
	\affiliation {Department of Physics, Engineering Physics, and Astronomy,
		Queen's University, Kingston, Ontario K7L 3N6, Canada}

%
%

	\author{Franco Nori}
\affiliation{Theoretical Quantum Physics Laboratory, RIKEN Cluster for Pioneering Research, Wako-shi, Saitama 351-0198, Japan} \affiliation{Physics Department, The University
	of Michigan, Ann Arbor, Michigan 48109-1040, USA}
%

%


\begin{abstract}

The quantum Rabi model is a widespread description for the coupling between a two-level system
and a quantized single mode of an electromagnetic resonator. Issues about this model's gauge invariance have been raised. These
issues become evident when the light-matter interaction reaches the so-called ultrastrong coupling regime. Recently, a modified quantum Rabi model able to provide gauge-invariant physical results in any interaction regime
{was introduced} [Nature Physics
{\bf 15}, 803 (2019)]. Here we provide an alternative derivation of this result, based on the implementation in two-state systems of the  gauge principle, which is the principle from which all the fundamental interactions in quantum field theory are derived.
The adopted procedure can be regarded as the two-site version of the general method used to implement the gauge principle in lattice gauge theories.
Applying this method, we also obtain the gauge-invariant quantum Rabi model for asymmetric two-state systems, and the multi-mode gauge-invariant quantum Rabi
model beyond the dipole approximation.

\end{abstract}

	\maketitle

\section{Introduction}

The ultrastrong  and deep-strong coupling (USC and DSC) between individual or collections of effective two-level systems (TLSs) and the electromagnetic field has been realized
in a variety of settings \cite{Kockum2018, Forn-Diaz2018}. In these 
{extreme} regimes of quantum
light-matter interaction,
the coupling strength becomes comparable  to (USC) or larger than (DSC) the transition frequencies of the system.

Recently, it has been argued that truncations of the atomic Hilbert space, to obtain a two-level description of the matter system, violate the gauge principle \cite{DeBernardis2018, Stokes2019,Stokes2020a}. Such violations become particularly
relevant in the USC and DSC regimes.
{In particular,
De Bernardis {\em et al.}~\cite{DeBernardis2018} shows that},  while in the electric dipole gauge, the two-level approximation can be performed as long as the Rabi frequency remains much smaller than the energies of all higher-lying levels, it can drastically
fail in the Coulomb gauge, even for systems with an extremely anharmonic spectrum.

The impact of the truncation of the Hilbert space of the matter system to only two states was  also studied 
{by Stokes and Nazir}~\cite{Stokes2019}, by introducing a one-parameter ($\alpha$) set of gauge transformations. 
The authors {found} that each  value of the
parameter produces a distinct quantum Rabi model (QRM), providing distinct physical predictions. {I}nvestigating a {\em matter} system with a lower anharmonicity (with respect to that considered in Ref.~\cite{DeBernardis2018}),
{they} use the gauge parameter $\alpha$  as a sort of fit parameter to determine the optimal QRM for a specific set of system parameters, by comparing the obtained $\alpha$-dependent lowest energy states and levels with the corresponding predictions of the  non-truncated gauge invariant model. The surprising result is that, according to this procedure, in several circumstances the {\em optimal} gauge is the so-called Jaynes-Cummings (JC) gauge, a gauge where the counter-rotating terms are automatically absent.

Recently, the source of gauge violation has been identified, and a general method for the derivation of light-matter Hamiltonians
in truncated Hilbert spaces able to produce gauge-invariant physical results has been developed{~\cite{DiStefano2019}}
(see also related work~\cite{Settineri2020, Savasta2020, Garziano2020}). 
This gauge invariance {was} achieved by compensating the non-localities introduced in the construction of the effective Hamiltonians. The resulting
quantum Rabi Hamiltonian in the Coulomb gauge differs significantly  from the standard one, but provides {exactly} the same
energy levels obtained by using the dipole gauge, as it should be, because physical observable quantities must be gauge invariant. 
A recent overview of these gauge issues in TLSs can be found in Ref.~\cite{LeBoite2020}.

Very recently, the validity of the gauge invariant {QRM} developed in Ref.~\cite{DiStefano2019} has been put into question {by Stokes and Nazir} \cite{Stokes2020a}.
Specifically, it is claimed that  the results  in Ref.~\cite{DiStefano2019} are not correct, and the truncation of the Hilbert space necessarily ruins gauge-invariance.

Here we present a detailed derivation of the results in Ref.~\cite{DiStefano2019} with an alternative, more direct and fundamental method.
{In}
our opinion, {this approach} demonstrates that the results in Ref.~\cite{DiStefano2019} [Eqs. (8) and (9) in particular] are indeed correct, and {moreover,}
represent the implementation in a fully consistent and physically meaningful way of the fundamental gauge principle in two-state systems. The derivation described here  can be regarded as the two-site version of the general method for {\em lattice gauge theories} \cite{Wiese2013}. These represent the most advanced and {commonly} used tool for describing gauge theories in the presence of {a truncated infinite-dimensional Hilbert space}. When a gauge theory is regularized
on the lattice, it is vital to maintain its invariance
under gauge transformations~\cite{Wiese2013}.
An analogous approach has been developed as early as 1933~\cite{Peirls1933} for the description of  tightly-bound electrons in a crystal in the presence of a slowly-varying magnetic vector potential (see, e.g., also {Refs.}~\cite{Luttinger1951, Hofstadter1976, Graf1995}).

Applying this method, we also obtain the multi-mode gauge-invariant {QRM} beyond the dipole approximation.

The derivation presented here in Sect.~\ref{GaugeTLS}, we {believe}, is already sufficient to eliminate any concerns about the validity of the results in Ref.~\cite{DiStefano2019}. However, in Sect.~\ref{Review}, we also provide  a reply to the key points raised {by Stokes and Nazir~\cite{Stokes2020a}. 

\section{The gauge principle}

In this section, we recall some fundamental concepts, which we will apply in the next sections.

In quantum field theory, the coupling of particles with fields is constructed in such a way
that the theory is invariant under a gauge transformation{~\cite{Maggiore2005}}. Here, we limit the {theoretical} model to consider  $U(1)$ invariance. This approach can be generalized to introduce non-abelian gauge theories \cite{Maggiore2005, Wiese2013}.

Let us consider the transformation of the particle field $\psi \to e^{i q \theta} \psi$.
This transformation represents a symmetry of the free action of the particle (e.g., the Dirac action) if $\theta$ is a constant, but we want to consider a generic function $\theta(x)$ 
({\it local} phase transformation).
However, the free Dirac action is not invariant under local phase transformations, because the factor  $e^{i q \theta(x)}$ does not commute with $\partial_\mu$.
At the same time, it is known that the action of the free electromagnetic field is invariant
under the gauge transformation:
\be\label{gaugeA}
A_\mu \to A_\mu - \partial_\mu {\theta} \, .
\ee
It is then possible to replace, in the action, the derivative $\partial_\mu$ with a {\em covariant derivative} of $\psi$ as
\be 
D_\mu \psi = \left( \partial_\mu + i q A_\mu \right) \psi\, ,
\ee
so that
\be
D_\mu \psi \to e^{i q \theta} D_\mu \psi\, ,
\ee
even when $\theta$ depends on $x$. 
It is now easy to construct a Lagrangian with a local $U(1)$ invariance.
It suffices to replace all derivatives $\partial_\mu$ with covariant derivatives $D_\mu$.

The same procedure, leading to the well-known minimal coupling replacement, can be applied to describe the interaction of a non-relativistic particle with the electromagnetic field. 
Considering a particle of mass {$m$} with a geometrical coordinate $x$ and a potential $V(x)$, the Hamiltonian of such a particle interacting with the electromagnetic field can be written as
\be\label{mcr}
\hat H_0^{\rm gi} = \frac{1}{2 m} \left[\hat p - q A (x) \right]^2  + V(x)\, ,
\ee
where $\hat p = - i d/dx$ is the momentum of the particle (here $\hbar =1$).
It turns out that the expectation values $\langle \psi | \hat H_0^{\rm gi} | \psi \rangle$ are invariant under local phase transformations,
\be\label{phasex}
\psi(x) \to e^{iq \theta(x)} \psi (x)\, ,
\ee
 thanks to the presence of the gauge field $A(x)$.

Note that the function of a continuous degree of freedom $\psi (x)$ lives in the infinite-dimensional space of all square-integrable functions, and the local phase transformation transforms a state vector in this space into a different vector in the same space. 
Finally, we observe that the total Hamiltonian, in addition to $\hat H_0^{\rm gi}$, includes the free Hamiltonian for the gauge field.


\section{Double-well systems  in the two-state limit}\label{doublewell}
The problem of a quantum-mechanical system whose
state is effectively restricted to a two-dimensional Hilbert
space is ubiquitous in physics and chemistry \cite{Leggett1987}. In the simplest
examples, the system simply possesses a degree of
freedom that can take only two values. For example, the
spin projection in the case of a nucleus of spin-$1/2$ or the polarization
in the case of a photon. Besides these
{\em intrinsically} two-state systems, a more common situation
is that the system has a continuous
degree of freedom $x$, for example, a geometrical coordinate, and a potential energy function
$V(x)$ depending on it, with two separate minima \cite{Leggett1987} (see \figref{fig1}). 
\begin{figure}[htpb] 
	\centering
	\includegraphics[width= 0.75 \linewidth]{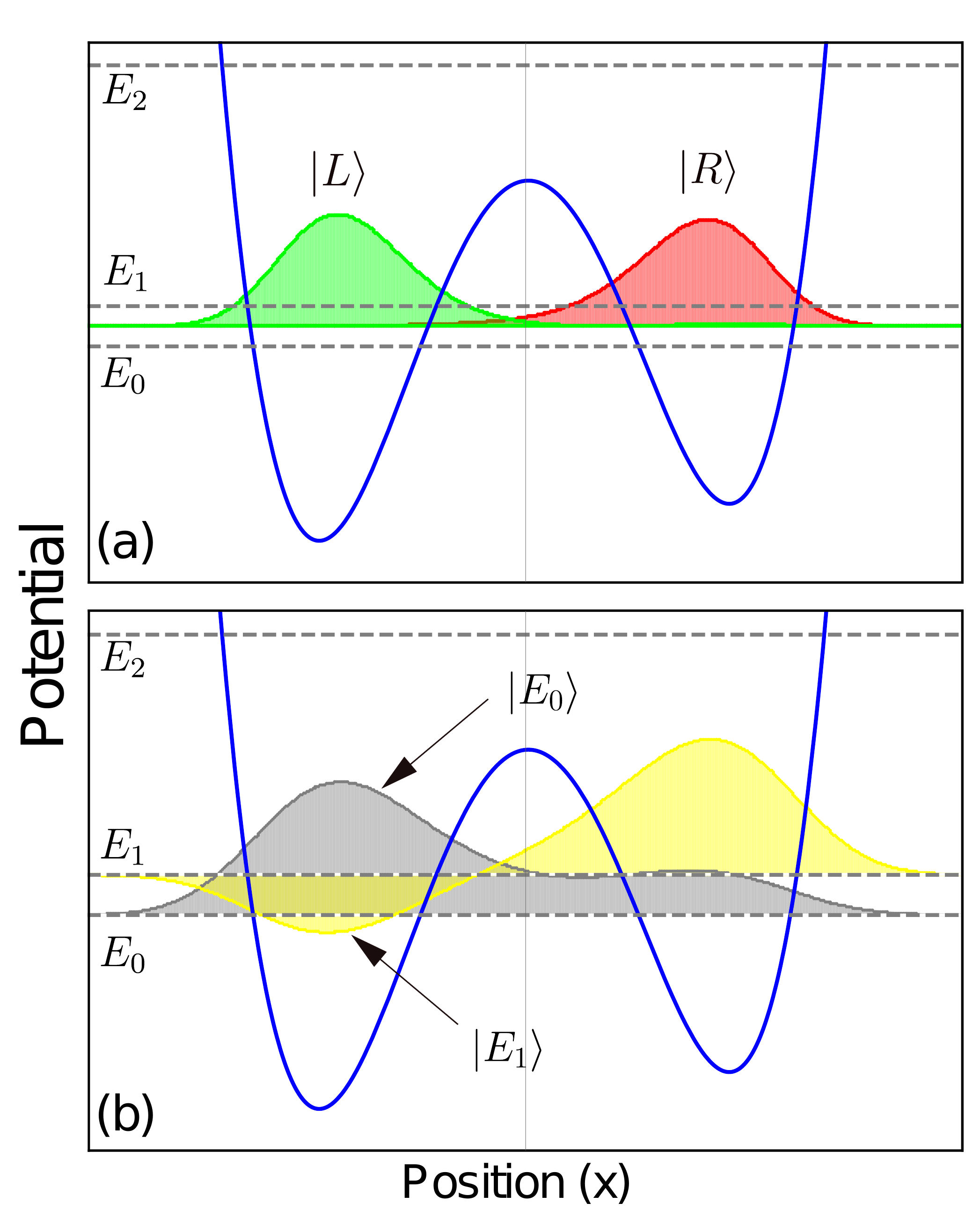} 
	\caption{{\bf A double-well system in the {\em two-state} limit}. {The symbols} $E_0$ and $E_1$ are the two lowest-energy levels, well separated in energy by the next higher energy level $E_2$. Panel (a) also shows the square modulus of the two wavefunctions localized in the well, obtained as linear combinations of the two lowest energy wavefunctions displayed in panel (b).}

	\label{fig1}
\end{figure}
 Let us assume that the barrier height $V$ is large enough that the system dynamics can be adequately described by a two-dimensional Hilbert space spanned by the two {\em ground} states in the two wells $| L \rangle$ and $| R \rangle$.

The motion in the two-dimensional Hilbert space can be adequately described by the simple Hamiltonian:
\be \label{H01}
\hat {\cal H}_0 =   \sum_{j= L,R} E_j | j \rangle \langle j  | - t \left( |R \rangle \langle L |+ {\rm h. c.}  \right)\, ,
\ee
where the tunneling coefficient is given by $t = \langle L | \hat H_0 | R \rangle$, and
\be\label{H0f}
\hat H_0 = \frac{\hat p^2}{2m} + V(x) 
\ee
is the usual system Hamiltonian.

If the potential is an even function of the geometrical coordinate,  $V(x) = V(-x)$ (see \figref{fig2}),  $E_L = E_R$, and we can fix $E_L = E_R = 0$. Introducing the  Pauli operator 
 $\hat \rho_x = |L \rangle \langle R |+ {\rm h.c.}$, we {obtain}
\be\label{H02}
\hat {\cal H}_0 = -t \hat \rho_x\, ,
\ee
whose eigenstates, delocalized in the two wells, are the well-known symmetric- and antisymmetric combinations (see \figref{fig2}b),
\bea
| S  \rangle &=& \frac{1}{\sqrt{2}} \left( |R \rangle + | L \rangle \right)\, , \nonumber \\
| A \rangle &=& \frac{1}{\sqrt{2}} \left( |R \rangle - | L \rangle \right)\, ,
\label{SA}\eea
with eigenvalues $E_{A,S} = \pm t$, so that $\Delta = E_A - E_S = 2t$, where we {assume} $t>0$.
The Hamiltonian in \eqref{H01} can be written in diagonal form as
\be \hat {\cal H}_0 = (\Delta/2) \hat \sigma_z\, ,
\ee
where $\hat \sigma_z = - \hat \rho_x = |A \rangle \langle A| - |S \rangle \langle S|$.
{Note, to distinguish between the different basis states for the operator representations,
we  use $\hat \sigma_i$
for the $\ket{A}{-}\ket{S}$ basis,
and $\hat \rho_i$ for the
$\ket{L}{-}\ket{R}$ basis.
Thus, for example, the diagonal $\hat \sigma_z$
operator becomes of nondiagonal
form in the $\ket{L}{-}\ket{R}$ basis.
}

It is worth noticing that this elementary analysis is not restricted to the case of a double-well potential. Analogous considerations can be carried out for systems with different potential shapes, displaying two (e.g., lowest energy) levels well separated in energy from the next higher level.
The wavefunctions $\psi_L(x) = \langle x |L \rangle$ and $\psi_R(x) = \langle x |L \rangle$ can be obtained from the symmetric and antisymmetric combinations of $\psi_S(x)$ and $\psi_A(x)$ (see \figref{fig1}), which can be obtained exactly as the two lowest energy eigenfunctions of the  Schr\"{o}dinger problem described by the Hamiltonian in \eqref{H0f}. The gap $\Delta = 2 t$ is obtained from the difference between the corresponding eigenvalues.
This two-state tunneling model is a well known formalism
to describe many realistic systems,
including the ammonia molecule, coupled quantum dots, and superconducting flux-qubits.

The case of a potential of the effective particle which does not display inversion symmetry can also be easily addressed. For example, an asymmetric double well potential, as shown in \figref{fig1}, can be considered.  
In this case,  \eqref{H01} can be expressed as 
\be\label{H0a1}
\hat {\cal H}_0 = \frac{\epsilon}{2} \hat \rho_z - \frac{\Delta}{2} \hat \rho_x\, .
\ee
The quantity $\epsilon$ is the {\em detuning} parameter,
that is, the difference in the ground-state energies of the states localized in the two wells in the absence of tunneling. The Hamiltonian in \eqref{H0a1} can be trivially diagonalized with eigenvalues $\pm \omega_q/2$, where $\omega_q = \sqrt{\Delta^2 + \epsilon^2}$.

\begin{figure}[htpb]  
	\centering
	\includegraphics[width= 0.75 \linewidth]{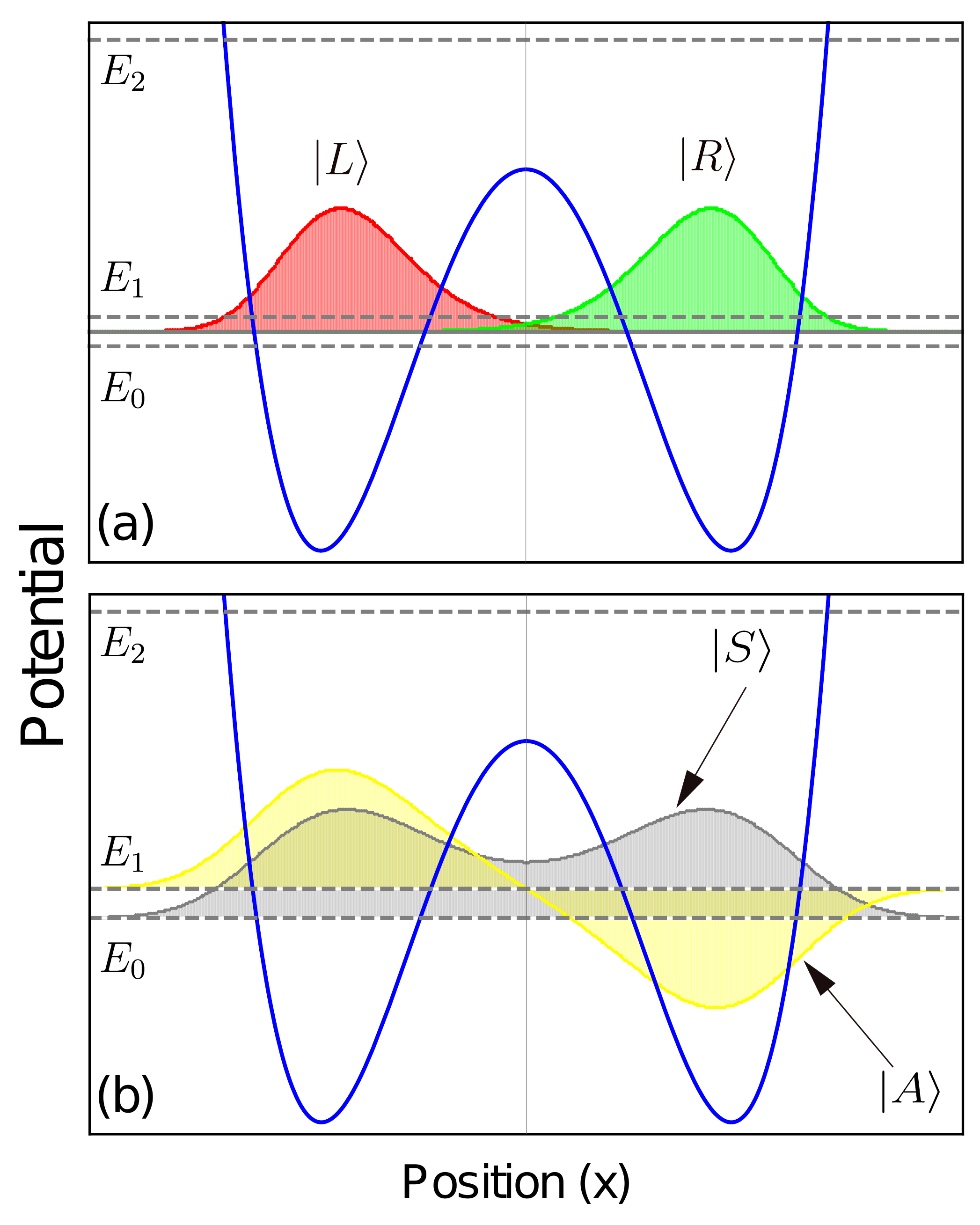}  
	\caption{{\bf A symmetric double-well system in the {\em two-state} limit}. {The symbols} $E_0$ and $E_1$ are the two lowest-energy levels, well separated in energy by the next higher energy level $E_2$. Panel (a) also shows the square modulus of the two wavefunctions localized in the well, obtained as symmetric and antysimmetric combinations of the two lowest energy wavefunctions displayed in panel (b).}

	\label{fig2}
\end{figure}

\section{The gauge principle in two-level systems}
\label{GaugeTLS}

The question arises if it is possible to {\em save the gauge principle} when, under the conditions described above, such a particle is adequately described by states confined {\em in a two-dimensional complex space}.
If we apply an arbitrary local phase transformation to, e.g., the wavefunction $\psi_A(x) = \langle x | A \rangle$: $\psi_A(x) \to \psi'_A(x) = e^{i q \theta(x) }\psi_A(x)$, it happens that, in general, $\psi'_A(x) \neq c_S \psi_S(x) + c_A \psi_A(x)$, where $c_A$ and $c_S$ are complex coefficients. Thus the general local phase transformation does not guarantee that the system can still be described as a two state system. According to this analysis, those works claiming {\em gauge non-invariance due to material truncation in ultrastrong-coupling QED} \cite{DeBernardis2018, Stokes2019, Stokes2020a} (we would say at any coupling strength, except negligible),  at first sight, might appear to be correct.

The direct consequence of this conclusion would be that two-level models, widespread in physics and chemistry, are too simple to implement their interaction with a gauge field, according to the general principle from which the fundamental interactions in physics are obtained. Since adding to the particle system description a few additional levels does  not change this point, the conclusion is even more dramatic. Moreover, according to {Stokes and Nazir}~\cite{Stokes2019, Stokes2020a},  this leads to several non-equivalent models of light-matter interactions {\em providing different physical results}. One might then claim the death of the gauge principle and of gauge invariance in truncated Hilbert spaces, namely in almost all cases where theoreticians try to provide quantitative predictions to be compared with experiments.

Our view is drastically different: we find that the breakdown of gauge invariance is the direct consequence of the {\em inconsistent} approach  of reducing the information (Hilbert space truncation) on the effective particle,  without accordingly reducing the information, by the same amount, on the phase $\theta(x)$ determining the transformation in \eqref{phasex}. In physics, the approximations must be done with care. These must be consistent.

We start by observing that the two-state system defined in \eqref{H01} {still has} {\em a geometric coordinate}, which however {\em can assume only two values}: $x_j$ (with $j= L, R$), {that} we can approximately identify with the position of the two minima of the double-well potential. More precisely, and {more} generally, they are:
\bea\label{xr}
x_R &=&  \langle R | x | R \rangle\, ,
\nonumber \\
x_L &=& \langle L | x | L \rangle\, .
\eea
Here, parity symmetry implies $x_L = - x_R$. In the following we will use the shorthand $\langle R | x | R \rangle = a/2$. Hence, the operator describing the geometric coordinate can be written as~\cite{Leggett1987}
${\cal X} = (a/2) \hat \rho_z$, where $\hat \rho_z \equiv | R \rangle \langle R| - | L \rangle \langle L |$.

We observe that the terms proportional to $t$ in the Hamiltonian in \eqref{H01} or \eqref{H02}, implies that these can be regarded as {\em nonlocal} Hamiltonians, i.e., with an effective potential depending on two distinct coordinates. Nonlocality here comes from the hopping term  $t = \langle R | \hat H_0 | L \rangle$, which is determined by the interplay of the kinetic energy term and of the potential energy in $\hat H_0$.

It is clear that the consistent and meaningful local gauge transformation corresponds to the following transformation
\be\label{phasec}
| \psi \rangle = c_L |L \rangle + c_R | R \rangle 
\to | \psi' \rangle = e^{i q \theta_L} c_L |L \rangle + e^{i q \theta_R}  c_R | R \rangle\, ,
\ee
where $| \psi \rangle$ is a generic state in the two-dimensional Hilbert space, and $\theta_j$ are arbitrary real valued parameters. 

It is easy to show that the expectation values of $\hat {\cal H}_0$ are not invariant under the {\em local } transformation in \eqref{phasec}.
They are only invariant under a uniform phase change: $| \psi \rangle \to e^{i q \theta}| \psi \rangle$.
 However, one can introduce in the Hamiltonian field-dependent factors that compensate the difference in the phase transformation from one point to the other. Specifically, following the general procedure of lattice gauge theory, we can {consider} the {\em parallel transporter} (a unitary {\it finite-dimensional} matrix), introduced by Kenneth Wilson \cite{Wilson1974,Lang2010,Wiese2013}, 
\be\label{W1}
U_{x_k +a,x_k} = \exp{\left[ i q \int_{x_k}^{x_k+a} dx\, A(x)   \right]}\, ,
\ee 
where $A(x)$ is the gauge field. After the gauge {transformation} of the field, $A'(x) = A(x) + d \theta / dx$, the transporter transforms as
\be\label{W2}
U'_{x_k +a,x_k} = e^{i q\, \theta(x_k+a)} U_{x_k +a,x_k} e^{-i q\, \theta(x_k)}\, .
\ee
This property can be used to implement gauge invariant Hamiltonians in two-state systems.

\subsection{Symmetric two-state systems}

Introducing {properly} the parallel transporter in \eqref{W1} into \eqref{H02}, we obtain a gauge-invariant two-level model:
\be\label{HL1}
\hat {\cal H}^{\rm gi}_0 = - t\, |R \rangle \langle L|\, U_{x_R, x_L} + {\rm h. c.}\, .
\ee
Gauge invariance can be directly verified: 
\bea
&\langle& \psi' | \left( |R \rangle \langle L|\, U'_{x_R, x_L} + {\rm h. c.} \right) |\phi' \rangle =  \nonumber \\
&\langle& \psi | \left( |R \rangle \langle L|\, U_{x_R, x_L} + {\rm h. c.} \right) |\phi \rangle\, ,\nonumber
\eea
where $| \psi \rangle$ and $| \phi \rangle$ are two generic states in the vector space spanned by $| L \rangle$ and $| R \rangle$.
By neglecting the spatial variations of the field potential $A(x)$ on the distance $a = x_R - x_L$, (dipole approximation). The Hamiltonian in \eqref{HL1} can be written as
\be\label{HLa}
\hat {\cal H}^{\rm gi}_0 = - t\, |R \rangle \langle L|\, e^{i q a A} + {\rm h. c.}\, .
\ee
Using \eqref{SA} and the Euler formula, it can be easily verified that the Hamiltonian in \eqref{HLa} can be expressed using the diagonal basis of $\hat {\cal H}_0$, as
\be\label{HL2}
\hat {\cal H}^{\rm gi}_0 =  \frac{\Delta}{2} \left[ \hat \sigma_z
 \cos{(q a  A)} + \hat \sigma_y \sin{(q a  A)}
 \right]\, ,
\ee
where $\hat \sigma_y = -i \left( | A \rangle \langle S| - |S \rangle \langle A| \right)$. Using \eqref{SA} and \eqref{xr},  {then} 
\be\label{d}
q a/2 = q \langle A | x | S \rangle\, . 
\ee
This {precisely} coincides with the transition matrix element of the dipole moment as in Ref. \cite{DiStefano2019}. 

Considering a quantized field $\hat A$, the total light-matter Hamiltonian also contains the free-field contribution:
\be\label{HL3}
	\hat {\cal H} =  \frac{\Delta}{2} \left[ \hat \sigma_z
	\cos{(q a \hat A)} + \hat \sigma_y \sin{(q a \hat A)}
	\right] 
	+ \hat H_{\rm ph}\, .
\ee

For the simplest case of a single-mode electromagnetic resonator,  the potential can be expanded in terms of the mode photon destruction and creation operators. Around $x=0$,  $\hat A = A_0 (\hat a + \hat a^\dag)$, where $A_0$ {(assumed real)} is the zero-point-fluctuation amplitude of the field in the spatial region spanned by the effective particle. We also have: $\hat H_{\rm ph} = \omega_{\rm ph} \hat a^\dag \hat a$, where $\omega_{\rm ph}$ is the resonance frequency of the mode. It can be useful to define the normalized coupling strength parameter \cite{DiStefano2019}
\be
\eta = q (a/2) A_0\, ,
\ee
so that \eqref{HL3} can be written as
\bea
\hat {\cal H} &=&  \frac{\Delta}{2} \left\{ \hat \sigma_z
\cos{[ 2 \eta (\hat a + \hat a^\dag) ]} + \hat \sigma_y \sin{[2 \eta (\hat a + \hat a^\dag)]}
\right\}\nonumber \\
&+& \omega_{\rm ph} \hat a^\dag \hat a\, .
\label{HL33}
\eea

Using  the relations $\hat \rho_z \equiv | R \rangle \langle R| - |L \rangle \langle L| = |A \rangle \langle S| + |S \rangle \langle A|\equiv \hat \sigma_x$,
the Hamiltonian in \eqref{HL1} can also be expressed as
\be\label{HL4}
\hat{\cal H} = \hat {\cal U} \hat {\cal H}_0 \hat {\cal U}^\dag\, ,
\ee
where 
\be\label{U0}
 \hat {\cal U} = \exp{(i q a \hat A \hat \sigma_x/2)}\, .
\ee
Equations (\ref{HL4}) and (\ref{U0}) coincide with Eqs. (8) and (9) of Ref.~\cite{DiStefano2019}, which represents the main results. 

It is also interesting to rewrite the {\em coordinate}-dependent phase transformation in \eqref{phasec} as the application of a unitary operator on the system states. Defining $\phi = (\theta_R + \theta_L)/2$ and $\theta = (\theta_R - \theta_L)/2$, \eqref{phasec} can be written as 
\be\label{phasec2}
| \psi \rangle \to | \psi' \rangle = e^{i q \phi} e^{i q \theta \hat \sigma_x} | \psi \rangle\, .
\ee
This shows that the {\em coordinate}-dependent phase change of a generic state of a {TLS} is equivalent to a global phase change, which produces no effect, plus {\em a rotation in the Bloch sphere}, which can be compensated by  introducing a gauge field as in \eqref{HL4}.
Notice {also} that {\eqref{phasec2} coincides} with the result presented in the first section of the  {Supplementary Information of Ref.~\cite{DiStefano2019}}, obtained with a different, but equivalent approach.
\vspace{0.4 cm}

In summary, the method {presented here} can be regarded as the two-site version (with the additional dipole approximation) of the general method for {\em lattice gauge theories} \cite{Wiese2013}, which represents the most advanced and {sophisticated} tool for describing gauge theories in the presence of truncation of infinite-dimensional Hilbert spaces.
These results eliminate any concern about the validity of the results presented in Ref.~\cite{DiStefano2019}, raised 
{by Stokes and Nazir}~\cite{Stokes2020a}.

We conclude this subsection by {noting} that \eqref{HL1} can be also used, without applying the dipole approximation, to obtain  the (multi-mode) {\em gauge-invariant quantum Rabi model beyond the dipole approximation}.	
Specifically, without applying the dipole approximation to \eqref{HL1}, after the same steps to obtain \eqref{HL33}, we obtain
\bea
	\hat {\cal H} &=&  \frac{\Delta}{2} \left[ \hat \sigma_z
	\cos{\left(q \int^{x_R}_{x_L}\! dx\, \hat A(x)\right)} \right. \nonumber \\ &+& \left. \hat \sigma_y \sin{\left(q \int^{x_R}_{x_L}\! dx\, \hat A(x)\right)}
	\right] 
	+ \hat H_{\rm ph}\, .
\label{Hbd}
\eea
One interesting consequence of this result is that it 
{introduces} {\em a natural cut-off} for the interaction of high energy modes of the electromagnetic field with a TLS. In particular, owing to cancellation effects in the integrals in \eqref{Hbd}, the resulting coupling strength between the TLS and the mode goes rapidly to zero when the mode wavelength becomes shorter than $a/2 = \langle A | x | S \rangle$.

It is worth noticing that  this derivation of the gauge-invariant QRM does not require the introduction of an externally controlled  two-site  {\em lattice} spacing, in contrast to general lattice gauge theories. In the present case, the effective spacing $a$ between the two sites is only determined by the transition matrix element of the position operator between the two lowest energy states of the effective particle, $a = 2 \langle A | x | S \rangle$, which in turn determines the dipole moment of the transition, $q a /2$.

\subsection{Asymmetric two-state systems}
The results in this section can be directly generalized to also address the case of a potential of the effective particle which does not display inversion symmetry.
It has been shown that the interaction (in the USC and DSC limit) of these TLSs (without inversion symmetry) with photons in resonators can lead to a number of  interesting phenomena \cite{Niemczyk2010, Ridolfo2012, Garziano2015, Garziano2016,Yoshihara2017,Kockum2017a, Stassi2017}.
In this
case, \eqref{H0a1} provides the bare TLS Hamiltonian.
Note that the first term in \eqref{H0a1} is not affected by the two-state local phase transformation in \eqref{phasec}, hence the gauge invariant version of \eqref{H0a1} can be written as
\be\label{HLa1}
\hat {\cal H}^{\rm gi}_0 = \frac{\epsilon}{2} \hat \rho_z -  \frac{\Delta}{2}\, \left(|R \rangle \langle L|\, U_{x_R, x_L} + {\rm h. c.}\right)\, ,
\ee
which, in the dipole approximation, reads:
\be\label{HLaa}
\hat {\cal H}^{\rm gi}_0 = \frac{\epsilon}{2} \hat \rho_z  - \frac{\Delta}{2}\, \left(|R \rangle \langle L|\, e^{i q a A} + {\rm h. c.}\right)\, .
\ee
{This} can be expressed as
\be\label{as1}
\hat {\cal H}^{\rm gi}_0 = \frac{\epsilon}{2} \hat \rho_z - \frac{\Delta}{2} \left[ \hat \rho_x \cos{(q a  A)} -  \hat \rho_y \sin{(q a  A)}\right]\,  ,
\ee
which can also be written in the more compact form
\be\label{mcrTLS}
\hat {\cal H}^{\rm gi}_0 = \hat {\cal U} \hat {\cal H}_0 \hat {\cal U}^\dag\, ,
\ee
where
\be\label{U}
\hat {\cal U} = \exp{\left[ i q a A \hat \rho_z /2\right]}\, .
\ee

{\em Equations~(\ref{mcrTLS}) and~(\ref{U})
represent the minimal coupling replacement for TLS, derived directly from the fundamental gauge principle.}

We observe that the operator $\hat {\cal X} = a \hat \rho_z / 2$ represents the geometrical-coordinate operator for the two-state system, with eigenvalues $\pm a/2$.
The Hamiltonian in \eqref{as1} can be directly generalized beyond the dipole approximation with the following replacement:
\be\label{bd}
a A \to \int_{-a/2}^{a/2} \!dx\, A(x)\, .
\ee

Considering a single-mode electromagnetic resonator, the 
total Hamiltonian becomes
\bea
\hat {\cal H} &=& \omega_{\rm ph} \hat a^\dag \hat a + \frac{\epsilon}{2} \hat \rho_z  \\ 
&-& \frac{\Delta}{2} \left\{ \hat \rho_x \cos{\left[2 \eta (\hat a + \hat a^\dag)\right]} -  \hat \rho_y \sin{\left[2 \eta (\hat a + \hat a^\dag)\right]}\right\}\, .
\nonumber
\eea

Since the operator $\hat {\cal X}$ is the position operator in the two-state space,  the unitary operator $\hat {\cal U}^\dag = \hat{\cal T}$ also corresponds to the  operator which implements the PZW unitary transformation \cite{Babiker1983}, leading to the dipole-gauge representation,
\bea
\hat {\cal H}_d &=& \hat {\cal U}^\dag \hat {\cal H} \hat {\cal U} = \omega_{\rm ph} \hat a^\dag \hat a  + \frac{\epsilon}{2} \hat \rho_z - \frac{\Delta}{2} \hat \rho_x \nonumber\\
&-& i \eta  \omega_{\rm ph} (\hat a - \hat a^\dag) \hat \rho_z +
\eta^2 \hat {\cal I}\, ,
\eea
where we used: $\hat \rho_z^2 = \hat {\cal I}$, {where} ${\cal I}$ is the identity operator for the two-state system.
Note that $\hat {\cal H}_d$ coincides  with the Hamiltonian describing a flux qubit interacting with an $LC$ oscillator \cite{Yoshihara2017}.

\section{Discussion}
\label{Review}

In Sect.~\ref{GaugeTLS}, we have derived from first principles the general QRM
for TLSs. The results presented in Sect.~\ref{GaugeTLS} exactly coincide with those obtained {by Di Stefano {\em et al.}~\cite{DiStefano2019} for symmetric TLSs and in the dipole approximation.
This derivation is already  sufficient  to eliminate  any  concern  about the validity of the results in 
Ref.~\cite{DiStefano2019},  recently raised {Stokes and Nazir~}\cite{Stokes2020a}.
However, {for  completeness,} here we 
{address} some of the specific criticisms  {that were raised}.

\subsection{The main issue raised by 
{Stokes and Nazir in Ref.~[5]}} }
\label{issueA}

The gauge-invariant approaches developed in Sect.~\ref{GaugeTLS}, and also in Ref.~\cite{DiStefano2019}, are in contrast with the point of view adopted by Stokes and Nazir \cite{Stokes2019,Stokes2020a}. According to them,
gauge  non-invariance is a necessary implication of the truncation of the Hilbert space of the material system.
Moreover, it was claimed \cite{Stokes2020a} that
the approach proposed in
Ref.~\cite{DiStefano2019} 
rests on an incorrect mathematical assertion and so does not resolve gauge non-invariance.

We have shown in Sect.~\ref{GaugeTLS}, not only  that the gauge-invariant QRM developed in Ref.~\cite{DiStefano2019} is correct, but also that 
it fits well in the  spirit of lattice gauge theories 
initiated by Kenneth Wilson \cite{Wiese2013}. Hence it is clear that the claims in Refs.~\cite{Stokes2020a,Stokes2019}
-- that gauge  non-invariance is a necessary implication of the truncation of the Hilbert space of the material system
-- {are} not correct.

Consistent with our approach, lattice gauge theories 
show that it is {\em vital} to maintain the gauge invariance of a theory after reducing the infinite amount of information associated to a continuous coordinate \cite{Wiese2013}, contrary to the claims of {Stokes and Nazir}~\cite{Stokes2019,Stokes2020a}.

In this section, we will also show that the method and the assumptions adopted in Ref.~\cite{DiStefano2019} to obtain a gauge invariant QRM and a general method to preserve gauge invariance in truncated Hilbert spaces are correct.

The apparent proof that the results in Ref.~\cite{DiStefano2019} 
rests on an incorrect mathematical assertion and does not resolve gauge non-invariance
 is presented in the section entitled
 {``Material truncation''}.

According to the authors{~\cite{Stokes2020a}}, the main issue is that
Ref.~\cite{DiStefano2019} 
{tacitly and incorrectly} equates 
\be
\hat P \exp{(i q x \hat A)} \hat P
= \exp{[iq (\hat P x \hat P) \hat A]} 
\, ,
\ee
where $\hat P$ is the projection operator for the TLS.
Since, they argue, $\hat P \neq \hat I$ (here $\hat I$ indicates the identity operator), in general, if $f (\hat O)$ is a nonlinear function of a {Hermitian}} operator $\hat O$, we have {$\hat P f(\hat O) \hat P \neq f(\hat P \hat O \hat P)$}.

Here, the key point from which all the criticisms descend, simply,  is  that Ref.~\cite{DiStefano2019} assumes  $\hat P = \hat I$, while, according to Stokes and Nazir \cite{Stokes2020a}, $\hat P \neq \hat I$. 
Specifically, the results in Ref.~\cite{DiStefano2019}, rely on the deliberate decision to treat the effective particle like a TLS, with its own identity. This is very clearly stated already below Eq.~(5). {Quoting directly} from Ref.~\cite{DiStefano2019}: ``$\hat P = |0 \rangle \langle 0| + |1 \rangle\langle 1|$ is the TLS identity
operator''.
{Clearly} $\hat P$ is an operator with a $2 \times 2 $ matrix representation. On the contrary, in Ref.~\cite{Stokes2020a},
{and using their different operator notation},
$P$ is an {\em infinite-dimensional} operator. From our perspective, the Hilbert space truncation occurs once and definitely. Moreover, once the matter system is described by a two-state system, any meaningful operator must act on this space and hence, it has a $2 \times 2$ matrix representation, and the properties of the identity operators can be legitimately  used. As a consequence, operations like, e.g., $\hat O^2 = (\hat P \hat O \hat P)^2$ are perfectly correct, in contrast to $\hat O^2 = (\hat P \hat O \hat P)^2$, with $\hat P$ defined as in Ref.~\cite{Stokes2020a}. 

References~\cite{Stokes2020a,Stokes2019} both consider the two-level approximation in a partial and (in our view) inconsistent manner, with operators that repeatedly can {\it bring the system in and out of the ``two-level" system}. Thus, strictly speaking, they do not have a rigorous two-level system, but one coupled to other external levels.  This unfortunate {\em mix-up} destroys gauge invariance as they show in their plots in Ref.~\cite{Stokes2019}.

In order to distinguish between the two different definitions of projection operators, from now on, we will indicate the projection operators as defined in 
Ref.~\cite{DiStefano2019} using calligraphic symbols.
Reference~\cite{DiStefano2019} starts {by using} (specifying it from the beginning) $\hat {\cal P} \equiv \hat {\cal I}$ already when deriving the 
dipole-gauge Rabi Hamiltonian in Eq.~(5), before presenting the main result~\cite{DiStefano2019}.
A consequence of this choice, is that the equivalence $\hat {\cal P} \hat U(x) \hat {\cal P} = \hat U (\hat {\cal P} x \hat {\cal P})$ is legitimate. This result is obtained by expanding $\hat U (x)$ in a Taylor series and then using for each term the relation $\hat {\cal P} x^n \hat {\cal P} = (\hat {\cal P} x \hat {\cal P})^n$, which can be easily obtained using the properties of identity operators. This procedure is described in detail at the beginning of the Section I of the Supplementary Material of Ref.~\cite{DiStefano2019} for a generic operator $\hat D(\theta) = e^{i q \theta(x)}$.

In summary, {Stokes and Nazir}~\cite{Stokes2020a} strongly criticizes the consequences of the choice $\hat {\cal P} \equiv \hat {\cal I}$. They explain that for a non-linear function $f$, [see Eq. (20)], $\hat P f(\hat O) \hat P \neq f(\hat P \hat O \hat P)$, which is of course correct if $\hat P$ is not the identity operator. However, we are {surprised to} see that, a few lines before Eq. (20), Ref.~\cite{Stokes2020a} seems to contradict itself, {by} using $\hat P f(\hat O) \hat P = f(\hat P \hat O \hat P)$ to derive their Eq.~(17).
 Specifically, following the procedure introduced in Ref.~\cite{Stokes2019}, 
 {Stokes and Nazir}~\cite{Stokes2020a} start from the {non}-truncated total Hamiltonian $H_\alpha$. Then, they apply the  projection operator $P$ to obtain their standard $\alpha$-gauge two-level model. However, they treat in a {\em different} way the free atomic Hamiltonian $H_m$ and the light-matter interaction term ${\cal V}_\alpha$. Specifically, they apply 
 the first  projection operator as $P H_m P$, but instead of applying the same procedure to the interaction term: ${\cal V}_\alpha \to P {\cal V}_\alpha P$, 
they use the non-equivalent (according to their definition of $P$) truncation:  ${\cal V}_\alpha(x,p) \to {\cal V}_\alpha (PxP, PpP)$.
Since ${\cal V}^\alpha (x, p)$ {is} a non-linear function of $x$ (it contains a quadratic term),   it is 
not at all clear why the authors
used ${\cal V}^\alpha (PxP, PpP)$, instead of $P {\cal V}^\alpha (x, p) P$.

The same procedure is also adopted and briefly described {by Stokes and Nazir} in Ref.~\cite{Stokes2019} (see the Methods section in particular). Below Eq.~(12), (which is the same, in a slightly different notation of Eq. (16) in Ref.~\cite{Stokes2020a}), the authors write:
\newline
\noindent
{``If the interaction Hamiltonian $V^\alpha$ is linear in ${\bf r}$ and ${\bf p}_\alpha$ then the two-level model Hamiltonian can also be written $H_2^\alpha = P^\alpha H P^\alpha$. This is not the case for $H$ in Eq.~(11) due to the ``${\hat {\bf d}}^2$" term, which demonstrates the availability of different methods for deriving truncated models. Here we adopt the approach most frequently encountered in the literature, and outline other methods in Supplementary Note 2.''}

In summary, not only do
{Stokes and Nazir}~\cite{Stokes2020a} use what a few lines below claim to be absolutely wrong, but, in a closely related work, the same authors also admit, after using this procedure,  that it is ``frequently encountered in the literature''.

\subsection{Gauge-ambiguities}
\label{GA}

{Stokes and Nazir} \cite{Stokes2020a} (see also Ref.~\cite{Stokes2019a}) also 
 point out that gauge-ambiguities are much broader than gauge non-invariance that results from an approximation. According to these references,
subsystem predictions vary significantly with the gauge relative to which the subsystems are defined
independent of model approximations.

Although it is true that features such as, e.g., the amount of light-matter entanglement and of bare excitations in the system eigenstates, are gauge-relative, in our opinion, this statement can be misleading and requires some comment.
In particular, we observe that, as described in detail in Ref.~\cite{Settineri2020}, the approach developed in Ref.~\cite{DiStefano2019} can be applied to remove gauge ambiguities  in  all the  experimentally  observable quantities  including the detectable light-matter  entanglement  in  the  ground  state  of  cavity-QED  systems,  even  in  the presence of Hilbert space truncation.  This is one of the main  results of Ref.~\cite{Settineri2020}. The main point here is that measurements (as, e.g., experimental clicks or transmission amplitudes) are numbers that do  not  care about our gauge discussions.  Therefore, if our approximations are applied consistently, as theoreticians, we should provide these numbers.  As theoreticians, we can play with different representations, but all of them must be consistent {and unambiguous}. Reference~\cite{Settineri2020} shows that this is the case even under extreme conditions, as  in  the  presence  of  deep  ultrastrong  light-matter  interactions  and/or  non-adiabatic ultrafast switches of the interaction. {These theories work well} even in the presence of relevant approximations, if these are carried out in a consistent manner.

\subsection{Other points raised by Ref.~[5]}

\noindent
{\bf {$\bullet$} Bloch sphere rotation}

 According to {Stokes and Nazir}~\cite{Stokes2020a},
the models actually analysed in Ref.~[4] are Bloch sphere rotations of the multipolar QRM.

As a matter of fact, 
the light-matter interaction is introduced, at a fundamental level, by invoking the gauge principle. This leads to the minimal coupling replacement. Then,   a gauge (unitary) transformation can be applied to obtain the multipolar gauge Hamiltonian, which acquires a simple form in the dipole approximation.
    In Sect. I of the Supplementary Material of Ref.~\cite{DiStefano2019}, it is explained how the results in Ref.~\cite{DiStefano2019} are {indeed} able to satisfy the gauge principle. Results in Sect.~\ref{GaugeTLS} confirm this result in a very clear and precise way. Hence, from a fundamental point of view, the opposite is true: The multipolar QRM works fine because it is a Bloch sphere rotation (as required by gauge invariance in TLSs) of the Hamiltonian in Eqs.~(8) and (9) of Ref.~\cite{DiStefano2019}. 
    
    As a final remark on this point, we observe that in the Coulomb gauge
    and using the QRM of Ref.~\cite{DiStefano2019}, the electric field operator can be expanded in terms of creation and destruction photon operators, as in the not-truncated model, and as in the free electromagnetic theory (in the absence of interactions). On the contrary, after the gauge transformation in the multipolar QRM, the electric field operator  also contains contributions from the atomic dipole, as a consequence of the PZW transformation. This is shown in detail in Ref.~\cite{Settineri2020}.
    This view is further confirmed by  \eqref{HL1} here, which can be used to derive the QRM beyond the dipole approximation in \eqref{Hbd}. A result that, {to} our knowledge, has never been obtained, so far within any gauge, including the multipolar gauge.

\vspace{0.25 cm}
\noindent
{\bf {$\bullet$} Non-equivalent models}
 
{Stokes and Nazir~\cite{Stokes2020a} state:},
``the idea of Ref. [4] (here Ref.~\cite{DiStefano2019}) to define two-level model gauge transformations as projections of gauge-fixing transformations does not resolve gauge non-invariance, because it does not produce equivalent models.''
{The authors  refuse our choice} to regard the operator $\hat {\cal P}$ as the identity operator for the two-level space, although it seems  that they also use it \cite{Stokes2019,Stokes2020a} (see {Sec.~}\ref{issueA}). Then, they realize \cite{Stokes2020a}  that refusing this implies that gauge non-invariance in truncated Hilbert spaces cannot be resolved. This is not surprising, since in Sect.~I of the Supplementary Material of Ref.~\cite{DiStefano2019}, it has been shown that the gauge principle is satisfied using repeatedly the properties of the identity operator.

{In our view, the} conclusion reached in Ref.~\cite{Stokes2020a} is a direct  consequence of an inconsistent choice attributed to Ref.~\cite{DiStefano2019}.
Sect.~\ref{GaugeTLS} here further confirms that only the choice $\hat P = \hat I$ can produce a gauge invariant theory in the spirit of lattice gauge theories.

\subsection{Some consequences of renouncing gauge invariance}

The consequence of the only possible choice, according to Refs.~\cite{Stokes2019,Stokes2020a},
is that the gauge principle cannot be implemented in truncated Hilbert spaces, and gauge transformations provide non-equivalent light-matter interaction models. Since this remains true beyond TLSs, and since almost every practical calculation involving field-matter interactions is carried out cutting the infinite amount of information provided by exact infinite-dimensional Hilbert spaces, {\em the unpleasant conclusion of
{Stokes and Nazir}~\cite{Stokes2020a,Stokes2019} is that  gauge invariance and the gauge principle do not work in most practical cases}. 
{Naturally, this} would be  a huge problem, not only in cavity QED, but also for a wide range of calculations, including the transport and optical properties of solids, especially in the presence of strong fields, and for the broad field of lattice gauge theories in quantum field theory and in condensed matter many-body quantum physics \cite{Wiese2013}.
As shown, e.g.,  in Refs.~\cite{Wilson1974,Lang2010, Peirls1933, Luttinger1951, Hofstadter1976, Graf1995},  \cite{Wiese2013} and references therein, {\em luckily} this is not the case  (see also Sect.~\ref{GaugeTLS}).

\section{On the existence of system-dependent {\em optimal} quantum Rabi models}
\label{NC}

In {Stokes and Nazir's} Ref.~\cite{Stokes2020a}, and also in
{their} Ref.~\cite{Stokes2019}, the authors correctly admit that the dipole gauge is optimal when the anharmonicity is high. However, a system with high anharmonicity is just, as also discussed in Ref.~\cite{DiStefano2019}, a system where the two level truncation can be safely performed, even in the presence of very high light-matter coupling strength. 

On the contrary, when the
 anharmonicity, $\mu$ (with $\mu = (\omega_{2,1}- \omega_{1,0})/{\omega_{10}}$) is of the same order or lower than the normalized light-matter interaction strength $\eta$ ({i.e., }$\mu \sim \eta$), the two level approximation becomes unreliable, because the detuning between the cavity frequency and the additional atomic transition frequencies becomes comparable with the coupling strength.
This trivial and well-known issue has been described in detail in Sect. V of the Supplementary Information of Ref.~\cite{DiStefano2019}.

In addition, if a strong positive detuning between the cavity-mode resonance frequency and the two-level transition frequency is considered,  the coupling of additional atomic transitions with the cavity photons becomes even more relevant,  and the two-level approximation becomes  pointless.
{Indeed, this} is the situation corresponding to a number of plots in Ref.~\cite{Stokes2019}.

Ref.~\cite{Stokes2020a},
citing Ref.~\cite{Stokes2019}, explains that the multipolar-gauge (what we call dipole-gauge) does not work when the material system is an harmonic oscillator. In Ref.~\cite{Stokes2019} the same concept is explained {as}: {``We show further that if the material system is a
	harmonic oscillator, then it is possible to derive a JCM (Jaymes Cummings model) that is
	necessarily more accurate than any derivable QRM (quantum Rabi model) for finding ground-state averages.''}

It is well-known that the spectra and the physical properties of a harmonic system constituted by two coupled harmonic oscillators is very far from those of a system constituted  of a two-level model interacting with a harmonic oscillator (QRM). Hence, the fact that the dipole-gauge QRM, which is an highly non-linear model is not able to describe harmonic oscillators is not surprising. 
However we are not able to catch the meaning of a highly nonlinear model, used as a fit to to reproduce only some very limited feature of the physics of two coupled harmonic oscillators
{(weak excitation limit)}.

We could elaborate in significantly more details and considerations, however, this is not the right place for a detailed analysis of the results in 
Ref.~\cite{Stokes2019}.

In the present work, we have shown how to apply the fundamental gauge principle to TLSs in order to derive a gauge-invariant QRM. Of course, this procedure works fine, until, taking also into account the interaction with the gauge field, the two-level approximation is meaningful. Naturally, if the detuning between the field and additional transitions becomes comparable with the coupling strength, these cannot be ignored anymore, and the two-level approximation is no more adequate.

\section{ Conclusions}

In contrast with the claims of 
{Stokes and Nazir} \cite{Stokes2020a}, we have shown in Sect.~\ref{GaugeTLS} of this work that the results presented in Ref.~\cite{DiStefano2019}, and also used in Ref.~\cite{Settineri2020}, are correct, deriving them with an alternative, more direct and fundamental method.
This derivation shows that the results in Ref.~\cite{DiStefano2019} are not only correct, but  they  constitute the  only  route (to our knowledge)  to  implement,  in  a  fully consistent  and  physically  meaningful  way,  the  fundamental  gauge  principle  in  truncated Hilbert spaces.
We have also extended the results in Ref.~\cite{DiStefano2019} to asymmetric two-state systems.

In addition, the method used here allowed us to obtain the gauge-invariant QRM beyond the dipole approximation, which is one of the main results of this work. Note that the problem of a quantum-mechanical system,  whose state is effectively restricted to a two-dimensional Hilbert space and which interacts with the electromagnetic field in various regimes, is ubiquitous in physics and chemistry \cite{Leggett1987}.
Hence, the availability of a general gauge-invariant  model describing this widespread physics is highly desirable.

The results in Sect.~\ref{GaugeTLS} shows that the results in Ref.~\cite{DiStefano2019}
{also fit} well in the great tradition of lattice gauge theories opened by Kenneth Wilson~\cite{Wiese2013}.
Lattice gauge theories
{constitute a powerful reference example where it is} possible and also {\em vital} to maintain the gauge invariance of a theory after reducing the infinite amount of information associated to a continuous coordinate \cite{Wiese2013}, contrary to the claims of Refs.~\cite{Stokes2019, Stokes2020a}.

A noteworthy feature of this derivation of the gauge-invariant {QRM} is that, in the present case, the two-site  {\em lattice} spacing is not externally controlled, in contrast to  general lattice gauge theories. Here, the effective spacing $a$ between the two sites is only determined by the transition matrix element of the position operator between the two lowest energy states of the effective particle, which in turn determines the dipole moment of the transition.

In Sect.~\ref{Review} we have also disproved the criticism {by Stokes and Nazir} \cite{ Stokes2020a}, claiming  that the results in  Ref.~\cite{DiStefano2019} rest on an incorrect mathematical assertion.

We conclude with some 
{key observations} from Sect.~\ref{GA}, where we pointed out that the analysis here and in Refs.~\cite{DiStefano2019, Settineri2020} also remove gauge ambiguities  in  the  experimentally  observable quantities,  including the detectable  light-matter  entanglement  in  the  ground  state  of  cavity-QED  systems,  even  in  the presence of Hilbert space truncation.  This is one of the main  results of Ref.~\cite{Settineri2020}. The key point is that {\em measurements} (as, e.g., experimental clicks or transmission amplitudes) {\em are data that do  not  care about  gauge representations}.  Therefore, if our approximations are applied consistently, as theoreticians, we should provide  numbers which are not affected by gauge transformations.  Of course, as theoreticians we can play with different representations, but all of them must be consistent. Reference~\cite{Settineri2020} shows that this is the case even under extreme conditions, as  in  the  presence  of  deep  ultrastrong  light-matter  interactions  and/or  non-adiabatic ultrafast switches of the interaction. All this works even in the presence of relevant approximations, if these are carried out in a consistent way.

If a theory is to be useful and meaningful, it should remove any ambiguities in the description of experimental data -- so a theory that is introduced to be ambiguous and not gauge invariant, ultimately is not very useful; so claiming such a theory is more correct is futile.

\begin{acknowledgements}

F.N. is supported in part by: NTT Research,
Army Research Office (ARO) (Grant No. W911NF-18-1-0358),
Japan Science and Technology Agency (JST)
(via 
the CREST Grant No. JPMJCR1676),
Japan Society for the Promotion of Science (JSPS) (via the KAKENHI Grant No. JP20H00134, and the grant JSPS-RFBR Grant No. JPJSBP120194828), and
the Grant No. FQXi-IAF19-06 from the Foundational Questions Institute Fund (FQXi),
a donor advised fund of the Silicon Valley Community Foundation.
SH acknowledges funding from 
the Canadian Foundation for Innovation, and
the Natural Sciences and Engineering Research Council of Canada.
%
S.S. acknowledges the Army Research Office (ARO)
(Grant No. W911NF1910065).
\end{acknowledgements}

\bibliography{refs}

\end{document}